# What is required for a post-growth model?

Van Eynde, Reo[1,*]; Dillman, Kevin J. [2]; Vogel, Jefim[1,3]; O'Neill, Daniel W. [1,4,*]

[1] UB School of Economics, Universitat de Barcelona, C/ de John Maynard Keynes 1-11, 08034, Barcelona, Spain
[2] Department of Environment and Natural Resources, School of Engineering and Natural Sciences, University of Iceland, 107 Reykjavík, Iceland
[3] Leeds University Business School, University of Leeds, Woodhouse, Leeds, UK
[4] Sustainability Research Institute,School of Earth and Environment, University of Leeds, Leeds, LS2 9JT, United Kingdom
* Corresponding author (r.vaneynde@ub.edu, oneil@ub.edu)

## Abstract

Post-growth has emerged as an umbrella concept for various sustainability visions that prioritise the pursuit of environmental sustainability, social equity, and human wellbeing, while questioning the pursuit of economic growth. Although there are increasing calls to include post-growth scenarios in high-level assessments, there is no coherent framework that specifies what is required to model post-growth. This article addresses this gap by: (1) identifying the minimum requirements for post-growth models, and (2) establishing a set of model elements for representing post-growth policy themes.

We survey post-growth modellers and draw on the post-growth literature to develop a framework of minimum requirements for post-growth models, integrating biophysical, economic, and social spheres and linking them to post-growth goals. Regarding the biophysical sphere, models should include resource use and pollution, environmental limits, and feedback mechanisms from the environment back to society, reflecting ecological embeddedness. Regarding the economic sphere, models should disaggregate households, incorporate limits to technological change and decoupling, include different types of government interventions, and calculate GDP or output endogenously. Regarding the social sphere, models should represent time use, material and non-material need satisfiers, and the affordability of essential goods and services. Specific policies and transformation scenarios require additional features, such as sectoral disaggregation or representation of the financial system.

Our framework aims to guide the development of models that can simulate both post-growth and pro-growth scenarios. Such models are needed to inform policymakers and stakeholders about the full range of options for pursuing sustainability, equity, and wellbeing.

# 1. Introduction

Humanity faces interrelated socio-ecological crises, in which multiple planetary boundaries are being transgressed (Richardson et al., 2023; Rockström et al., 2024), while billions of people remain deprived of their essential needs (Fanning and Raworth, 2025). Growing inequality within and between nations is resulting in abundance for a few while perpetuating scarcity for the many (Brand-Correa et al., 2022). Some strands of post-growth literature see the capitalist political-economic system, with its continued pursuit of economic growth even in high-income countries, as a key driver of these environmental and social challenges (Schmelzer et al., 2022). In contrast to claims that economic growth could continue in a sustainable manner (Bowen and Hepburn, 2013; Drummond et al., 2021), post-growth scholars question the feasibility and desirability of the continued pursuit of GDP growth in wealthy nations (Hickel et al., 2022; Jackson and Victor, 2019).

Post-growth research argues that tackling the environmental crises requires a stabilisation or reduction of global material and energy use and explores how social and environmental goals can be achieved without economic growth (Jackson and Victor, 2019). We define post-growth as an umbrella concept encompassing sustainability visions that prioritise environmental sustainability, social equity, and human wellbeing over economic growth. Following this definition, post-growth includes approaches such as degrowth, steady-state economics, the wellbeing economy, and Doughnut economics (Hardt and O'Neill, 2017; Kallis et al., 2025).

Macroeconomic models are an important tool for informing economic and political decisions and thus play a pivotal role in addressing the intertwined social and environmental challenges. They can be used to simulate how relevant variables evolve under different scenarios and policies. Here, scenarios are understood as sets of model parameters that affect the model behaviour, and policies are understood as intentional interventions in real-world leverage points.

However, several researchers have questioned the capabilities of mainstream economic models to address these challenges (Fullbrook, 2004; Keen, 2021; Stern, 2013; Stern et al., 2022). A first major critique of these models from a post-growth perspective is their treatment of economies as isolated from the environment and society (Daly and Farley, 2011), which underrepresents the systemic impacts of environmental damages (Keen, 2021; Spash and Hache, 2022). A second critique is that model studies tend to use GDP per capita as a proxy for human wellbeing – despite numerous studies highlighting that GDP is a poor measure of human wellbeing and that increases in GDP beyond a certain level of material wealth display little or even negative association with human wellbeing (Diener et al., 2010; Easterlin, 1995; Fanning and O'Neill, 2019). Lastly, mainstream models have been criticised for optimising a single goal, rather than considering multiple incommensurable and non-substitutable goals, such as environmental sustainability, social equity, and human wellbeing (Costanza et al., 2025; O'Neill, 2020).

These critiques highlight the need for *post-growth models* and *post-growth scenarios*. By a post-growth model, we mean a model that can simulate and evaluate scenarios in a manner that is consistent with core post-growth ideas in the literature. By a post-growth scenario, we mean a scenario that represents key post-growth policies, such as resource caps, progressive taxation, or universal basic services. In recent years, the post-growth literature has seen an increase in modelling studies (Edwards et al., 2025; Lauer et al., 2025), while the Intergovernmental Panel on Climate Change (IPCC) and ecological economists have called for the inclusion of post-growth scenarios in high-level assessment reports (IPCC, 2023a; Keyßer and Lenzen, 2021; Slameršak et al., 2024).

Although important contributions have been made related to post-growth modelling, we identify two research gaps. First, despite proposals to improve the representation of post-growth scenarios in models (e.g. Kikstra et al., 2024; Van Eynde et al., 2024; Victor and Jackson, 2020), a framework that specifies requirements for post-growth models is lacking in the literature. Second, it is not clear what minimum set of elements is required to model specific post-growth policies. This study aims to address these gaps by introducing a framework that provides the minimum requirements for models to simulate post-growth policies and scenarios and evaluate their outcomes against post-growth goals.

Understanding the requirements for post-growth models is crucial to facilitate the integration of post-growth scenarios into high-level assessment reports and to inform critical economic and political decisions. Furthermore, this understanding is required to avoid the misrepresentation of post-growth scenarios and policies. In particular, post-growth models should be able to differentiate between a post-growth transition and a recession, as the two are often conflated. Although we propose a framework for post-growth models, we argue that these models should be able to simulate not only post-growth scenarios. On the contrary, they should simulate and compare different types of scenarios to provide better guidance on how to achieve environmental sustainability, social equity, and human wellbeing.

The remainder of this article is structured as follows. Section 2 introduces the concept of post-growth, its goals and policy themes, and applications to modelling. Section 3 discusses the methods underpinning our analysis, in particular our survey of post-growth modellers. Section 4 discusses the results from the survey. Section 5 proposes a conceptual framework for post-growth models based on these results. Section 6 discusses tensions in modelling post-growth scenarios, limitations to modelling, and limitations of this research, while Section 7 concludes.

# 2. Post-growth and modelling

There is no consensus on the use of the term "post-growth", it is sometimes used interchangeably with degrowth (Vincent and Brandellero, 2023), while others argue for a separation of the concepts degrowth and post-growth (Likaj et al., 2022; Wiedmann et al., 2020). We use post-growth as an umbrella concept for growth-critical sustainability visions that share a focus on environmental sustainability, social equity, and human wellbeing, in line with various authors in the literature (e.g. Gerber and Raina, 2018; Hardt and O'Neill, 2017; Kallis et al., 2025).

Examples of these visions include steady-state economics, degrowth, the Doughnut of social and planetary boundaries, and the wellbeing economy. *Steady-state economics* aims to stabilise resource use within environmental limits and move beyond GDP as society's measure of progress (Daly, 1977). *Degrowth* calls for a radical rethinking of society and the replacement of the global capitalist economic system with one that is governed according to the principles of (environmental) justice, sufficiency, and conviviality (Schneider et al., 2010). The *Doughnut of social and planetary boundaries* envisions sustainability as the satisfaction of the essential needs of the global population while staying within planetary boundaries (Raworth, 2017a). The *wellbeing economy* proposes a narrative that is focused on the principles of human wellbeing, fairness, participation, and sustainability (Coscieme et al., 2019; McCartney et al., 2023).

## 2.1. Post-growth goals

Following principles of strong sustainability (Daly, 1977; Neumayer, 2013), the three goals of post-growth — environmental sustainability, social equity, and human wellbeing — should be considered incommensurable and non-substitutable (Gough, 2017; O’Neill et al., 2018). It is crucial to explore and model interdependencies and potential trade-offs between these goals (Lamb, 2016; Millward-Hopkins and Oswald, 2023; O’Neill et al., 2018; Vogel et al., 2021).

The goal of **environmental sustainability** reflects the observation that key Earth systems underpin human wellbeing and societal functioning but that their functioning is threatened by anthropogenic impacts (Brondizio et al., 2019; IPCC, 2023b; Rockström et al., 2024). The post-growth literature stresses the importance of indicators that reflect the pressure on major Earth systems with respect to critical thresholds such as planetary boundaries. To assess a country’s compatibility with planetary boundaries, its impacts can be compared to a downscaled national boundary, which requires some operationalisation of equity or national fair-shares (Hickel et al., 2022; O’Neill et al., 2018; Vogel and Hickel, 2023). More localised environmental impacts such as air pollution or soil depletion are also crucial but may require a different approach (Van Eynde et al., 2024).

Another key concern of post-growth is **social equity** in domains such as income, wealth, opportunity or wellbeing dimensions – across countries but also within countries across social strata such as class, gender, race, or age. Low levels of inequality are considered an important societal goal (Fanning et al., 2020; O’Neill et al., 2018; Raworth, 2017a), and most people report strong preferences for low income inequality (Millward-Hopkins and Oswald, 2021). Levels of income inequality also strongly impact social outcomes such as access to sanitation or sufficient nourishment (Vogel et al., 2021). Furthermore, international power dynamics and ecologically unequal exchange are important dimensions to consider (Dorninger et al., 2021; Hickel et al., 2024; Lang, 2024).

The post-growth literature conceptualises **human wellbeing** primarily in relation to the satisfaction of human needs, which is considered a prerequisite for social participation and human flourishing (Büchs and Koch, 2017; Gough, 2017; Koch et al., 2017). Human needs are seen as universal, non-substitutable, and satiable, whereas need satisfiers (the goods and services used to satisfy needs) are context-dependent and somewhat substitutable (Doyal and Gough, 1991; Max-Neef, 1991). Both objective indicators (e.g. life expectancy) and subjective indicators (e.g. life satisfaction) are important for measuring human wellbeing (Costanza et al., 2009; O’Neill et al., 2018). Post-growth modelling should thus seek to represent the satisfaction of human needs, adequately incorporating their drivers and satiability characteristics. GDP is not seen as an adequate indicator of wellbeing (Costanza et al., 2025, 2014; Stiglitz et al., 2009).

## 2.2. Post-growth policy themes

The post-growth literature contains a wide variety of policy proposals. Several scholars have reviewed this literature and classified the policies into themes that cover policies with similar goals or focuses (Cosme et al., 2017; Fitzpatrick et al., 2022; Hardt and O’Neill, 2017). Drawing on their work, we summarise the post-growth policy space in eight high-level policy themes (Table 1). This classification includes most of the policy themes discussed in the previously mentioned literature, except for some themes that can only partially be represented by macroeconomic variables (e.g. democracy and culture). The mapping of the policies from the literature to our policy themes can be found in the Supplementary Materials 1 (Table S1).

**Table 1. The list of post-growth policy themes considered in this study.**

| Policy theme | Description |
| --- | --- |
| *Goal-oriented policies* | |
| Reduce human impact on Earth systems | Policies to reduce the environmental pressure of societies, e.g. through caps or taxes on resource use, transition to renewables, biodiversity restoration, protected areas, or rights of nature. |
| Reduce inequality | Policies to reduce inequality through the equitable (re)distribution of income, wealth, or access to resources. |
| Promote wellbeing and needs satisfaction | Policies that promote wellbeing, through providing access to goods and services, and ensuring decent living standards for everyone. |
| *Transformation policies* | |
| Change consumption patterns | Policies to change consumption behaviour, with the goal of promoting sufficiency of consumption and wellbeing. |
| Change production patterns | Policies to change production patterns to align them better with social and environmental goals. |
| Transform the financial system | Policies to transform financial and monetary systems to ensure economic stability, enable shifts towards socially necessary and ecologically sustainable production and consumption, or to promote financial democracy. |
| Rethink international relations | Policies to promote just trade practices and just international relations, which support environmental and social goals. |
| Transform work | Policies to restructure the organisation of paid and unpaid work, to align these with environmental sustainability and human wellbeing. |

These policy themes can be grouped into two categories. The first category consists of the policy themes that directly aim to improve one of the three post-growth goals. We call this category "goal-oriented policies", and it includes policies that reduce human impact on Earth systems, reduce inequality, and promote wellbeing and needs satisfaction.

The second category — which we call "transformation policies" — are policy themes that aim to structurally change the way our socioeconomic systems operate and to overcome growth dependencies. These policies are seen as instrumental in achieving the three main post-growth goals, but they diverge further from policies proposed in mainstream sustainability science. Each of the transformation policies may affect more than one of the three main post-growth goals simultaneously.

## 2.3. Post-growth modelling

In this article we focus on models with a national/regional scale, or a global scale with national/regional resolution. While we mainly consider relatively large models that are calibrated to real-world data and provide a detailed granular representation of environment–society–economy systems, we acknowledge that smaller or more theoretical models are also relevant, for example for studying very high-level trends and scenarios (Crespo Cuaresma, 2017; Leimbach et al., 2017) or for testing innovative features (e.g. Dasgupta, 2021). In recent years, post-growth scenario modelling has seen a strong increase, with a systematic literature review identifying 75 modelling studies that are related to degrowth and/or post-growth (Lauer et al., 2025). Furthermore, Edwards et al. (2025) reviewed 15 post-growth scenarios that were used in modelling studies and identify five core types of policies to achieve the key post-growth goals.

To date, post-growth modelling has mainly been taken up in two strands of literature. The field of ecological macroeconomics has engaged the most with building models that are specifically tailored to studying post-growth scenarios (Hardt and O'Neill, 2017). Arguably, the first model that produced low growth scenarios was the World3 model from the *Limits to Growth* report (Meadows et al., 1972). More recently, ecological macroeconomists have made promising steps towards modelling post-growth scenarios, with models such as *LowGrow SFC* (Jackson and Victor, 2020), *EUROGREEN* (D'Alessandro et al., 2020), *MEDEAS* (Capellán-Pérez et al., 2020; Nieto et al., 2020), and *DEFINE* (Dafermos and Nikolaidi, 2022a, 2022b). For a more detailed overview of the current state of the art in post-growth modelling studies, we refer the reader to Lauer et al. (2025) and Edwards et al. (2025).

Post-growth scenarios have also been explored with integrated assessment models (IAMs), although their original purpose was not the simulation of post-growth scenarios. These models are designed to explore the interactions between the environment, society, and the economic system (IPCC, 2023a). They have been influential in steering climate policy but have been criticised, amongst other things, for their optimistic assumptions about the decoupling of GDP growth from environmental pressures, which obscures the role of economic growth in driving up emissions (Hickel and Kallis, 2020; Larkin et al., 2018). Several calls have been made to include degrowth/post-growth scenarios in these mainstream modelling approaches and to rethink existing models (Keyßer and Lenzen, 2021; Slameršak et al., 2024). IAM community members have called for realistic IAM representations of socio-technical transitions (Geels et al., 2017; 2016) and for transition narratives that depend on behavioural or broader regime changes (Trutnevyte et al., 2019; van Sluisveld et al., 2020). In recent years, some efforts have been made to implement post-growth scenarios in existing IAMs, such as the International Futures model (Moyer, 2023) and MESSAGEix (Kikstra et al., 2024; Li et al., 2023). Furthermore, the IAM community has proposed Sustainable Development Pathways to address some of the critiques of Shared Socioeconomic Pathways (Hernandez et al., 2024; Soergel et al., 2024).

In both strands of literature, various proposals have been made for how to better represent post-growth scenarios in models. For instance, calls have been made for the inclusion of non-monetary economic relations (Hardt and O'Neill, 2017), a better representation of the financial sector (Victor and Jackson, 2020), the distinction between essential and luxury goods (Kikstra et al., 2024), a more detailed representation of provisioning systems and social and environmental indicators (Van Eynde et al., 2024), and the interdependencies between the Global North and Global South (Lauer et al., 2025). Although each of these proposals captures important aspects from the post-growth literature, a framework that provides the requirements for post-growth models remains lacking.

# 3. Methods

To address this gap, we developed a survey to query researchers working on post-growth modelling. Drawing on the literature, we created a list of model elements that are relevant from a post-growth perspective (Table 2). In the survey, we then asked respondents which elements are relevant for modelling each post-growth policy theme (Table 1). Building on these results, we created a framework of minimum requirements for post-growth models, and recommendations for modelling specific policy themes. The remainder of this section discusses the creation of the survey, the sample of respondents, and how the survey results were analysed.

First, we created a list of relevant model elements (a functional set of model variables) for post-growth modelling and grouped them into thematic clusters (Table 2). This list was compiled by drawing on the post-growth literature, macroeconomic theory, and conversations with experts. Note that some model elements could arguably be placed in more than one category (e.g. employment comes out of the interaction of households with firms).

**Table 2. Model elements relevant for post-growth modelling, grouped according to thematic clusters.**

| Cluster | Model elements |
|---|---|
| Needs satisfaction | Material needs satisfiers (e.g. food, housing)<br>Non-material needs satisfiers (e.g. social inclusion, peace/safety)<br>Affordability of essential goods<br>Distinction between essential and luxury goods |
| Biophysical sphere | Resource flows<br>Energy flows<br>Anthropogenic impacts on ecosystems (e.g. pollution, deforestation)<br>Limits on physical resource use/extraction<br>Ecological impacts on human systems |
| Modes of provisioning | For-profit organisations (e.g. firms)<br>Not-for-profit organisations (e.g. NGOs)<br>Government (e.g. public services)<br>Non-institutional provisioning (e.g. community, households, dark economy) |
| Households | Differentiation/disaggregation of household types (e.g. by gender or skill level)<br>Consumption decisions (quantities, types of products/services)<br>Employment<br>Time use<br>Income<br>Wealth |
| Firms | Differentiation/disaggregation of economic sectors<br>Production decisions (quantities, types of products/services)<br>Price setting<br>Investment decisions<br>Resource/pollution intensities of production<br>Technological change |
| Government | Redistribution (e.g. welfare transfers and taxation)<br>Price controls (e.g. rent caps)<br>Environmental protection/regulation (e.g. taxation, conservation)<br>Steering investments (e.g. government investment, subsidies)<br>Government bonds and interest rates |
| Central bank | Monetary policy (e.g. interest rates, quantitative easing)<br>Financial regulation (e.g. minimum reserve requirements, Basel III, capital adequacy) |
| Financial sector | Price and direction of credit<br>Disaggregation of financial assets (e.g. loans, bonds, equity)<br>Disaggregation of financial firms (e.g. banks, pension funds, insurance companies)<br>Financial stability (e.g. balance sheet solvency, liquidity)<br>Ownership structures and control of financial institutions |
| International interactions | International trade in monetary terms<br>Resources and pollution embodied in trade<br>Other international financial flows (e.g. debt/interest payments, aid)<br>Exchange rates |

Second, we created a survey that asked respondents to identify which of the model elements from Table 2 are needed to model the policy themes from Table 1, with a focus on the elements that are directly related to a specific theme. We further asked respondents to identify relevant model elements that they thought were missing from the survey. We also included questions on respondent experience with modelling. Before being used, the survey was revised based on the feedback from two external experts.

Third, we sent the survey to 209 researchers working in the field of post-growth modelling. Specifically, we contacted people in the Post-Growth Modelling Community mailing list (which was established at the 2024 Degrowth and European Society for Ecological Economics conference in Pontevedra), authors whose modelling work was included in the review by Lauer et al. (2025), and additional respondents through snowballing. We received 38 responses (a response rate of 18%). On average, the respondents had 12 years of modelling experience, with a range of 0 to 36 years.

Fourth, we calculated the percentage of respondents that indicated a direct relationship between a model element and a policy theme for each element–theme combination. In our analysis, a value of 0% signifies a complete consensus that there is no direct relationship, while a value of 100% signifies a complete consensus that there is a direct relationship. We used the Chi-squared test to assess whether the percentage for each element–theme combination was significantly different from 50% (the midpoint between perfect consensus for and perfect consensus against a direct relationship). For our sample size, the result was that an agreement among 76.3% of respondents was required to indicate a statistically significant consensus ($p < 0.05$) that a given model element is required to model a given policy theme.

Supplementary Materials 1 contains additional information about (1) policy themes and model elements mentioned in the post-growth literature, (2) the survey that we used, and (3) further results from the survey that are not mentioned in the main manuscript.

# 4. Survey results

This section discusses the results from the survey with regards to the goal-oriented policies (Section 4.1), the transformation policies (Section 4.2), and overall relevance of each element (Section 4.3). The results in this section correspond to the full sample of respondents. In the Supplementary Materials we provide the results considering only the experts per policy theme. In what follows, we present policy themes and element clusters in *italics*, while individual model elements are in plain text.

Overall, the results show that to represent the three goal-oriented policy themes simultaneously, a model should include most elements from the clusters *biophysical sphere*, *needs satisfaction*, and *households*.

Except for the clusters *financial sector* and *central bank,* some elements from the other clusters are also required. Regarding the transformation policy themes, the results suggest that the required elements strongly depend on the specific theme. Furthermore, the themes *change production patterns* and *transform the financial system* seem to require the inclusion of a larger number of elements than the other transformation policy themes. Finally, across all policy themes, respondents showed the strongest consensus on including the following elements: government and for-profit modes of provisioning, material needs, income, and government steering of investments.

## 4.1. Goal-oriented policies

The respondents showed a consensus on the inclusion of six to eight elements for each of the respective goal-oriented policy themes (Figure 1). For the theme *reduce human impact on Earth systems*, the respondents agreed that most of the elements from the cluster *biophysical sphere* should be included, namely resource flows and limits, energy flows, and anthropogenic ecosystem impacts. Although the respondents did not significantly agree on the inclusion of ecological impacts, its value (74%) lies very close to the significance threshold. From the other clusters, the identified elements were technological change, resource intensities, environmental protection, and embodied resources and pollution in trade.

Regarding the theme *reduce inequality*, respondents identified various relevant elements from the *households* cluster, namely disaggregated households, income, and wealth. Other relevant elements identified by respondents include the affordability of essential goods, government provisioning, redistribution, and price controls.

For the policy theme *promote wellbeing and need satisfaction*, respondents agreed most strongly on the elements from the cluster *needs satisfaction*, more specifically on material and non-material needs, and the affordability of essential goods. The agreement about including the distinction between essential and luxury goods again fell two percentage points short of the significance threshold. Other elements with a significant consensus were government provisioning, a mechanism for redistribution, and time use.

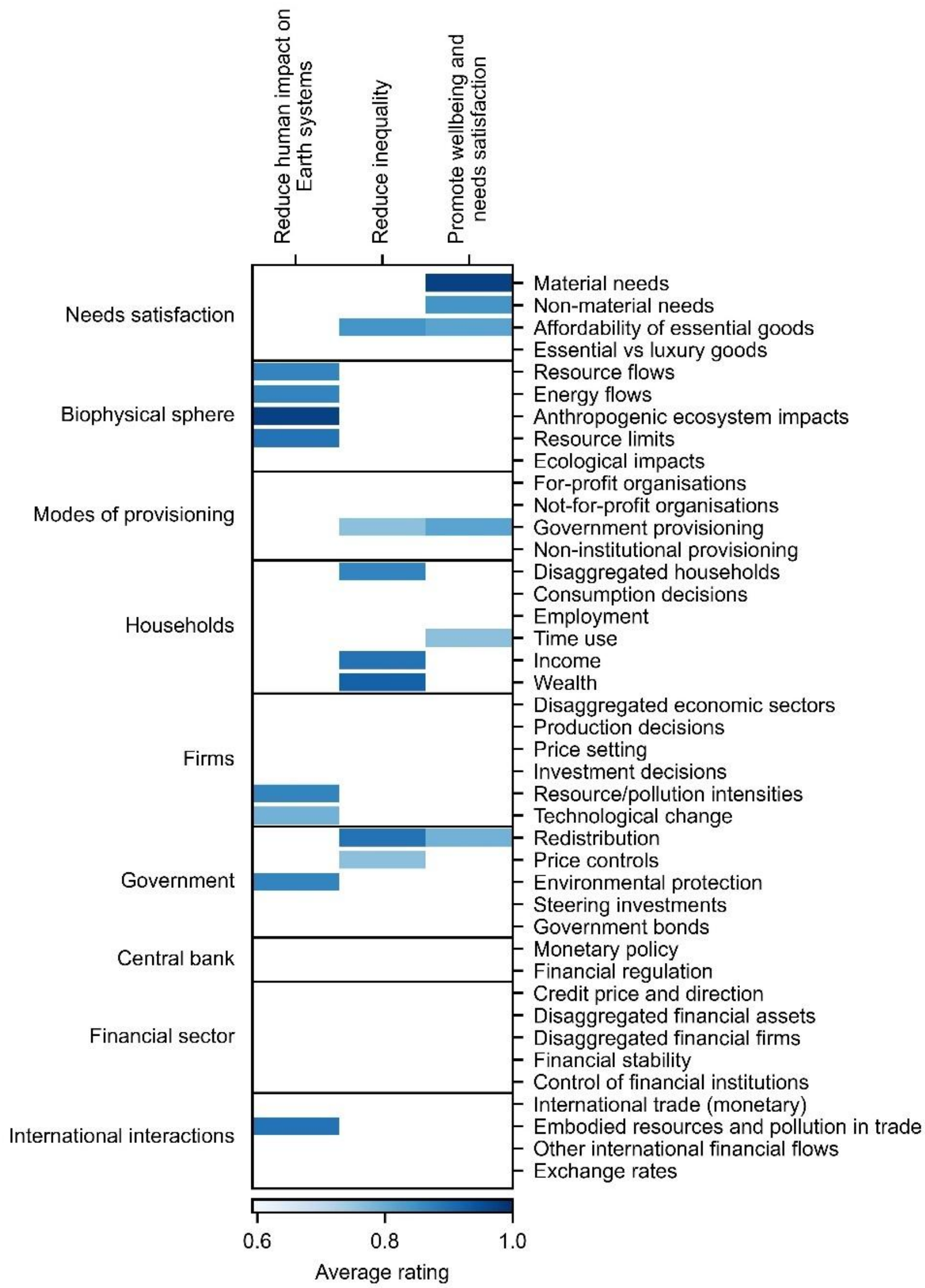

**Figure 1. The degree of consensus on whether a given model element is required to model the three goal-oriented post-growth policy themes.** The three goal-oriented policies are shown along the top, element clusters on the left, and model elements on the right. A coloured cell indicates a statistically significant consensus among respondents that this element is required to model a specific policy theme. The colour scale indicates the fraction of respondents that marked a model element as directly relevant for modelling a given policy theme.

Although each theme is most strongly linked to one specific cluster, the results indicate that adequately representing the goal-oriented policy themes requires including elements from a variety of clusters. To simultaneously cover the three policy themes, respondents suggest that several elements from the clusters *needs satisfaction*, *biophysical sphere, households*, and *government* should be included. Furthermore, the results suggest that a few elements from *modes of provisioning, firms*, and *international interactions* are required. The only clusters for which the majority of the respondents did not see a direct relationship with the goal-oriented policy themes were *financial sector* and *central bank*.

## 4.2. Transformation policies

To model the theme *change consumption patterns*, respondents indicated that the relevant elements are material needs, essential vs luxury goods, resource flows, consumption decisions, and household income (Figure 2). For the theme *change production patterns*, respondents agreed upon all but one element from *firms*, in addition to resource and energy flows, for-profit and government provisioning, and the steering of investments by the government. Regarding *transform the financial system*, respondents selected all elements in the clusters *central bank* and *financial sector*. Furthermore, we found consensus for the inclusion of government bonds, other international financial flows, and exchange rates. The theme *rethink international relations* was linked to all elements in the cluster *international interactions*, and not to any other cluster. The theme *transform work* was related to the elements employment, time use, and income, and to technological change.

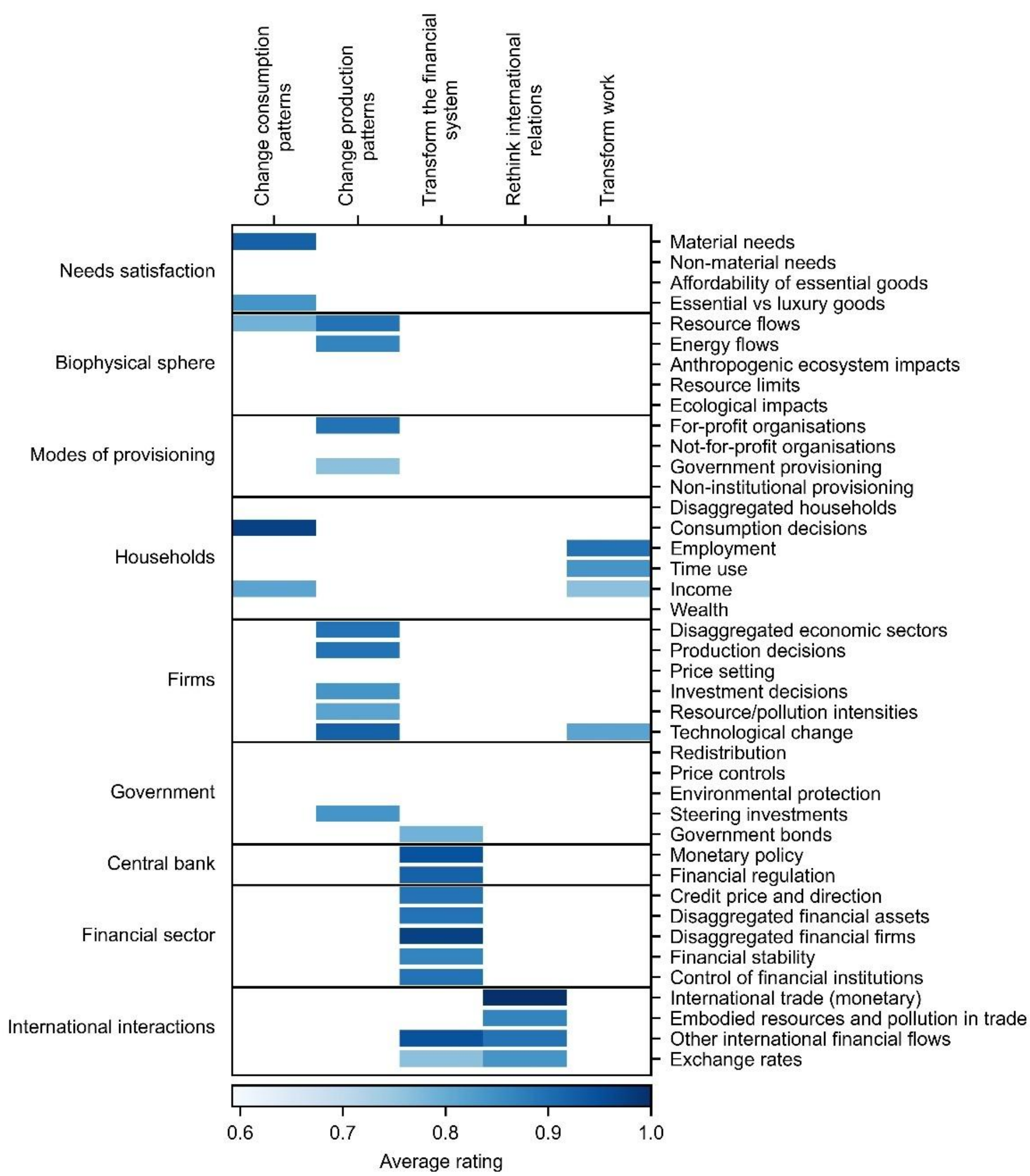

**Figure 2. The degree of consensus on whether a given model element is required to model l the post-growth transformation policy themes.** Policy themes are shown along the top, element clusters on the left, and model

elements on the right. A coloured cell indicates a statistically significant consensus that this element is required to model a specific policy theme. The colour scale indicates the fraction of respondents that marked a model element as directly relevant for modelling a given policy theme.

The results suggest that the theme *change production patterns* requires a broad scope of elements from various clusters, and that the theme *transform the financial system* requires a detailed representation of the financial sector and some elements related to international interactions. In contrast, the other three transformation policy themes seem to have a narrower scope, as respondents agreed only on a smaller set of elements which are mainly concentrated in one or two element clusters. Furthermore, each cluster is related to at most two transformation policy themes, suggesting that whether model elements should be included depends on the type of policies one wants to simulate or the research questions one wants to address.

## 4.3. Overall relevance of model elements

To assess the overall consensus on the importance of the elements, we analysed the average score that each element received across all policy themes (Figure 3). Note that although many elements surpass the significance threshold of 76.3% for certain policy themes, no element surpasses this threshold across all themes. Therefore, the average score across themes lies below this threshold. A first insight from these results is that the different forms of provisioning are seen as important elements across policy themes, with government and for-profit provisioning at the top of the list, and non-institutional provisioning in the top quartile. The respondents see an important role for government provisioning when modelling the themes *promote wellbeing and needs satisfaction*, *change production patterns*, and *reduce inequality*. For-profit provisioning is seen as relevant mainly in relation to *transform production patterns* and t*ransform work*.

Other consistently selected elements are material need satisfiers, income, and steering investments. The modelling of material need satisfiers seems to be a core element for modelling post-growth, as for each of the goal-oriented policy themes more than 70% of the respondents agreed on its importance. Similarly, income has a score of above 70% for four policy themes. Although steering investments is also ranked highly in the aggregate results, this ranking is mainly shaped by the strong consensus on its inclusion for the policy theme *change production patterns*.

A last observation is that most of the lowest-ranked elements are related to financial aspects of an economy. They are either part of the element clusters *financial sector* or *central bank*, or refer to financial aspects of other clusters, such as government bonds or financial regulation. Most respondents only see a direct connection between these elements and the theme *transform the financial system*.

The respondents also identified three elements as important for post-growth policy themes but missing from the survey (Supplementary Materials 1, Table S3). The first element is the labour market, which respondents saw as important for the policy theme *transform work*. Second, respondents mentioned ecosystem functioning and integrity, which would require a stronger engagement with the field of ecology (e.g. Gonzalez et al., 2020; Symstad et al., 2003). The third element refers to the non-monetary side of international relations, such as international power dynamics, competition, and institutions, which could shed light on how international institutions maintain unequal power relations or give rise to growth imperatives in the Global South (Lang, 2024; Musthaq, 2021).

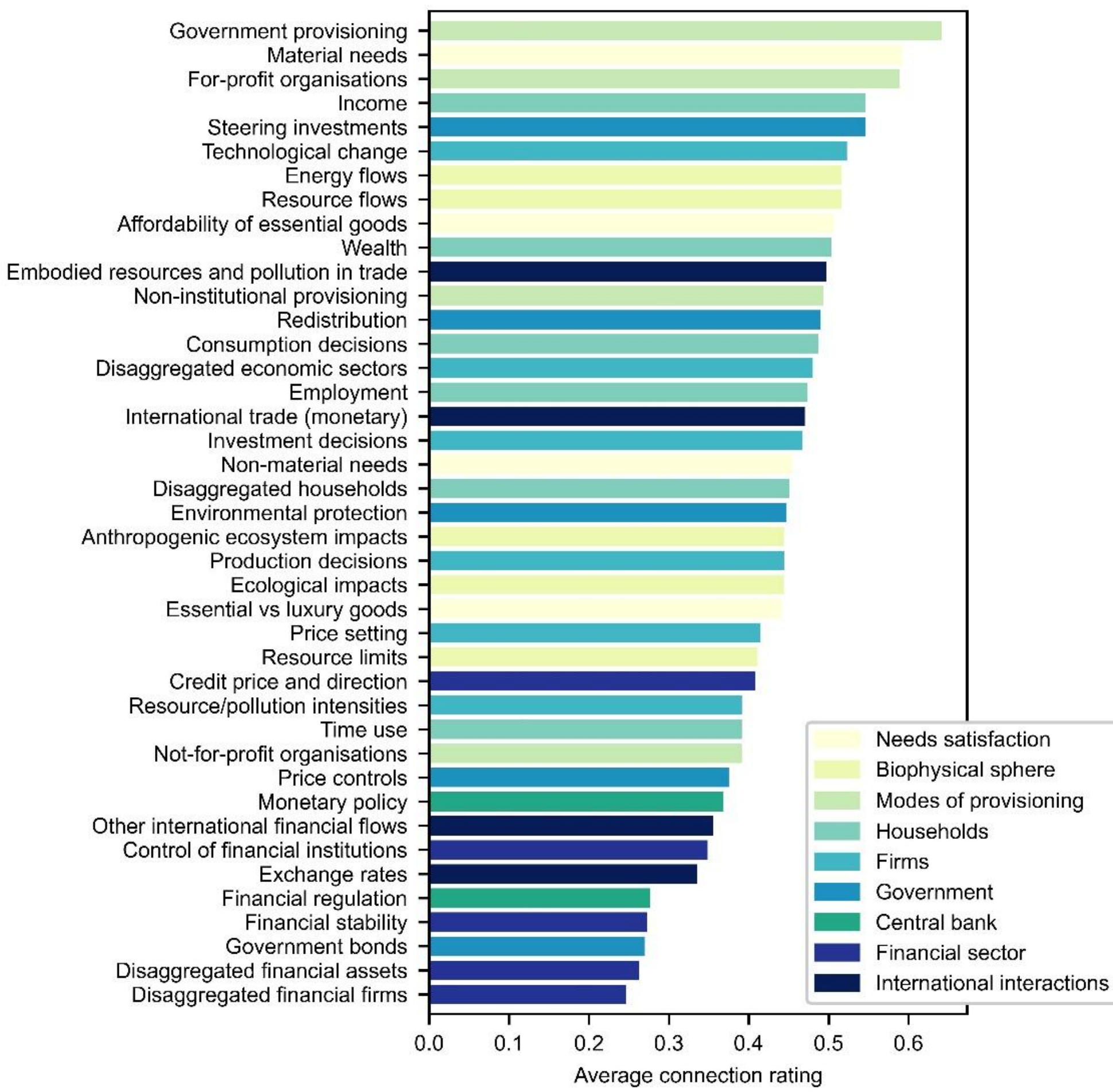

**Figure 3. The consensus that a given model element is important for modelling post-growth, as average across policy themes.** Model elements are shown on the left, and colour-coded according to which cluster they belong to. A value of 1 means that all respondents marked a direct relationship between the element and all eight of the policy themes. A value of 0 means that no respondent marked a direct relationship between the element and any of the policy themes.

# 5. Requirements for a post-growth model

In this section, we outline what is required for a post-growth model, based on the results from the previous section and drawing on the post-growth literature. We first propose a conceptual framework that outlines minimum model requirements for modelling post-growth scenarios in line with core post-growth ideas from the literature (Section 5.1). We then discuss additional elements that may be required to model specific policies or address specific research questions (Section 5.2).

## 5.1. A conceptual framework

For a model to be capable of modelling post-growth, we propose as a minimum requirement that it should assess scenario outcomes in terms of the three main goals of environmental sustainability, social equity, and human wellbeing. By assessing these three goals, post-growth models should be able to distinguish between post-growth scenarios and recessions. We argue that, in addition to adhering to core ideas established in the post-growth literature, post-growth models should also include the elements that were linked to the goal-oriented policies in our survey results (Figure 1). Based on these elements, we propose a conceptual framework for how these elements can be integrated in a model (Figure 4).

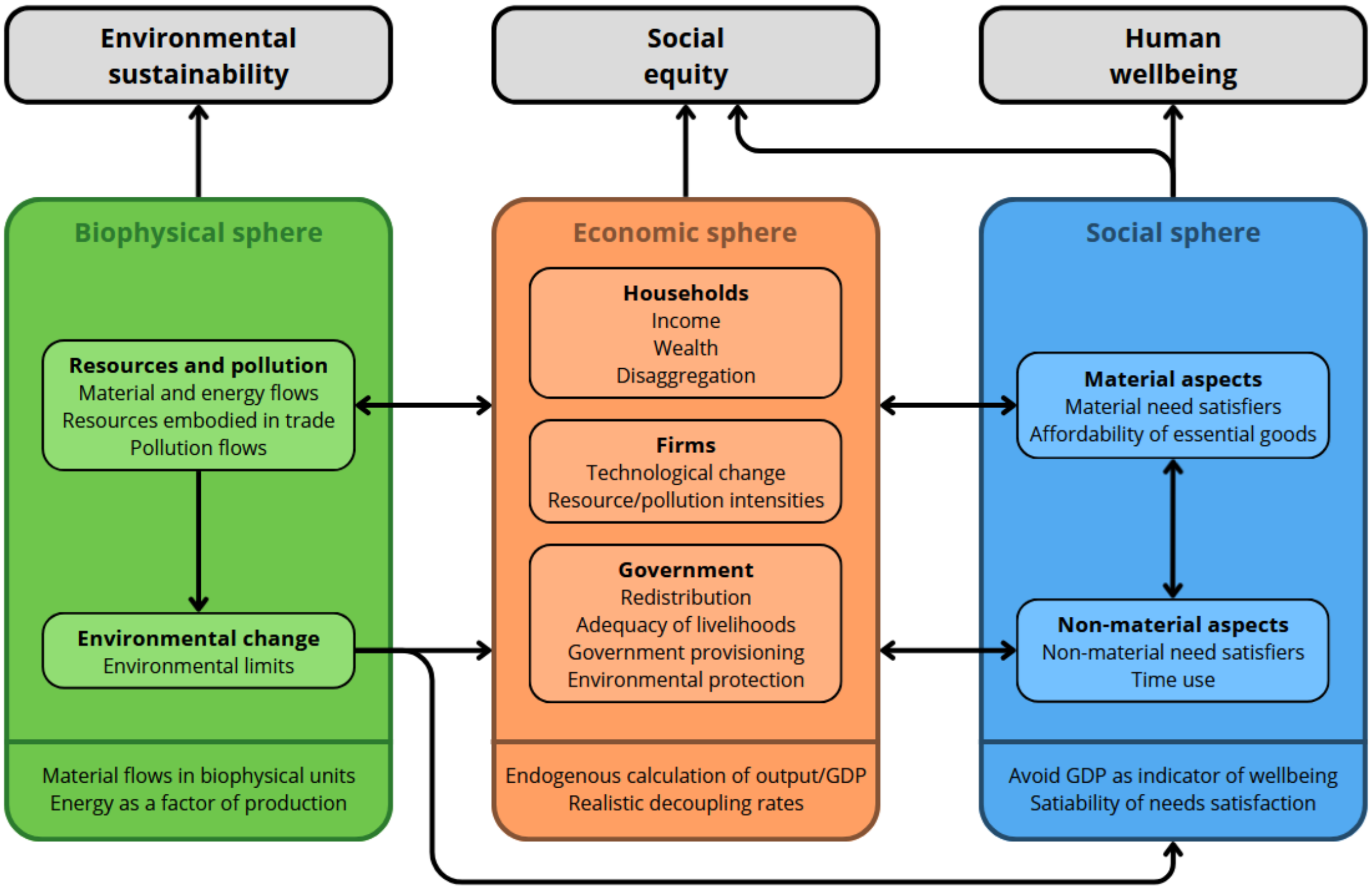

**Figure 4. A framework showing the requirements for a post-growth model.** The three coloured blocks represent the biophysical, economic, and social spheres. The grey blocks represent the three post-growth goals, while the arrows from the spheres to the goals indicate which spheres mainly affect which goals. The blocks in a lighter shade within each sphere represents a concept (in bold) and the corresponding model elements (regular font). Arrows between concepts and/or spheres indicate interdependencies between these. The statements at the bottom of the spheres describes additional considerations that are relevant for that sphere.

Our framework consists of three interrelated parts, namely the biophysical, economic, and social spheres. As such, the framework has a broader system boundary than mainstream economic models, which primarily focus on monetary variables (Hardt and O'Neill, 2017). The framework is consistent with the idea of the embedded economy (Daly, 1991; Raworth, 2017b), a core concept in ecological economics and the post-growth literature. A post-growth model should include the biophysical flows between the biophysical and economic spheres, and feedback mechanisms from the biophysical sphere back to the economic and social spheres. The economic sphere should be represented at a minimum through the interaction of three main agents: households, firms, and the government. In line with the literature on social provisioning and provisioning systems, activity in the economic sphere drives social outcomes, which should be evaluated in terms of the satisfaction of human needs, considering both material and non-material need satisfiers (Fanning et al., 2020; Jo and Todorova, 2017). We now discuss each of these three spheres in more detail.

### 5.1.1. The biophysical sphere

The principle of embeddedness has three implications for post-growth models. First, an economy requires material and energy inputs and creates pollution (Georgescu-Roegen, 1971), which a model should represent in biophysical units to express the dependence of the socioeconomic system on the biosphere (Victor and Jackson, 2020). Models should also represent the interdependencies between different environmental dimensions, such as how climate mitigation through renewable energy deployment requires non-negligible amounts of materials and land (Nieto et al., 2020). Second, energy (or exergy) should be treated as a factor of production and should thus be included in production functions (Gonçalves et al., 2024; Keen et al., 2019; Santos et al., 2024). Furthermore, there should be constraints on production and consumption due to resource availabilities (Baranzini et al., 2015; Dafermos et al., 2017). Third, embeddedness implies that models should represent how human activity affects crucial life-supporting Earth systems, and how changes to these systems have effects on the economic and social spheres. Environmental impacts should be compared to relevant sustainability and equity thresholds to capture trade-offs and the complexity of balancing various environmental constraints.

### 5.1.2. The economic sphere

In the economic sphere, households, firms, and the government should be represented. The inclusion of households is important because they provide labour, save and invest, and are the beneficiaries of the provisioning process. To obtain an accurate representation of inequality (e.g. income, wealth), models should disaggregate households, for example based on employment status, skill level, gender and age. Disaggregation across demographic dimensions can highlight structural inequalities, such as racial inequalities, gender pay gaps or unequal care burdens (Blecker and Braunstein, 2022; Cieplinski et al., 2023). Furthermore, population dynamics affect economic dynamics (e.g. through labour supply, dependency ratios, or pension benefits payments) and should therefore be considered.

Regarding the representation of firms, it is important to represent resource and pollution intensities to evaluate the impacts of production on ecosystems. Second, technological change is relevant because it affects how capital and labour productivity — as well as resource/pollution intensities — evolve over time. Post-growth research warns that decoupling of economic activity from environmental impacts is insufficient to reconcile continued economic growth with environmental targets, and that mainstream scenarios often make highly optimistic assumptions about decoupling rates (Hickel and Kallis, 2020; Keyßer and Lenzen, 2021; Vogel and Hickel, 2023). The efficacy of decoupling strategies is limited by (1) limits to thermodynamic efficiency (Keen et al., 2019), (2) limits to the pace of technological change (Geels et al., 2017; Vogel and Hickel, 2023), and (3) rebound effects that tend to counteract potential energy savings from energy efficiency gains (Brockway et al., 2021). Models should reflect such limits to the efficacy of decoupling strategies for reducing environmental pressures.

In the post-growth literature, the government is seen as a critical actor for achieving the three post-growth goals (Cosme et al., 2017; D’Alisa and Kallis, 2020). However, a conceptualisation of government intervention that goes beyond government spending is often lacking (Corlet Walker et al., 2021). Based on the survey results, we suggest three key roles for governments. First, the government can implement redistributive policies to reduce inequality between social groups, which may in turn increase the public acceptability of post-growth policy proposals (Paulson and Büchs, 2022). Second, the government can ensure the adequacy of livelihoods, for instance through price controls on essential goods and services produced in the market, or through the free provisioning of essential goods and services (Coote and Percy, 2020; Gough, 2019; Vogel et al., 2024). Third, governments can protect the environment through the implementation of policies such as resource caps, environmental taxes, or biodiversity protection (Fitzpatrick et al., 2022).

Three additional considerations related to economic dynamics are key. First, economic growth occurs only under specific political-economic conditions (Schmelzer, 2015). Therefore, a model should not assume economic growth *a priori*, but should calculate output or GDP endogenously, not least to assess how societies can move beyond economic growth (Edwards et al., 2025). Although GDP is a poor measure of human wellbeing, the representation of total economic activity is relevant as it relates strongly to environmental pressures, and there is value in being able to make comparisons to mainstream models.

Second, potential growth dependencies and growth imperatives should not be assumed to be universal but should be examined by modelling actual causal structures and moderating factors (Stratford, 2023; Vogel et al., 2024; Wiman, 2024). One salient example is how GDP (or income per capita) is often used as a driver of social outcomes, which embeds a growth dependency in models (Van Eynde et al., 2024). Variables should only be modelled as functions of GDP where this is theoretically and empirically supported.

Lastly, post-growth policy proposals include both supply-side and demand-side changes, which cannot be fully represented in purely supply-driven or purely demand-driven models. Ideally, post-growth modelling should therefore reflect interdependencies between the supply and demand side and enable modelling both types of interventions.

### 5.1.3. The social sphere

We argue that post-growth models should seek to represent both material and non-material aspects of the social sphere. Material aspects include material need satisfiers and the affordability of essential goods and services. We propose three considerations for modelling material need satisfiers. First, due to the satiability of needs, need satisfaction returns on increases in consumption rapidly diminish beyond a certain level, which is often best described through saturation curves (Steinberger and Roberts, 2010; Vogel et al., 2021). Second, modelling multiple material needs requires a disaggregation into different economic sectors or provisioning systems that each relate to the provisioning of specific need satisfiers (Hickel et al., 2022). Third, when modelling need satisfaction as a function of resource use, it is important to account for key provisioning factors (Vogel et al., 2021).

An additional material aspect relates to the affordability of, and access to, essential goods and services. Modelling affordability and access requires specifying types and quantities of goods and services that meet minimum thresholds for needs satisfaction (e.g. Davis et al., 2024; McElroy and O'Neill, 2025; Rao and Min, 2018). In economies where essential goods and services must be purchased, needs can only be satisfied when people have access to and are able to afford these essentials. Through the inclusion of disposable income and prices of essential goods and services, models can assess the adequacy of livelihoods and how it is affected by policies (Mayrhofer and Wiese, 2020; Vogel et al., 2024).

Regarding non-material aspects, the first required element is non-material need satisfiers, i.e. non-material ways of meeting human needs, which are especially relevant for needs such as social participation or social relationships. Despite being non-material, people may sometimes try to satisfy or substitute for these needs in materialistic ways, such as through consumption (Jackson and Marks, 1999). In existing macroeconomic models, non-material need satisfiers have only seen limited uptake (Van Eynde et al., 2024). Although there is a clear relationship between certain economic variables and non-material aspects of wellbeing (Helliwell et al., 2024; Roffia et al., 2023), other drivers of need satisfaction are harder to represent endogenously in a macroeconomic framework. Nevertheless, these drivers can be included as exogenous variables and could be changed through scenario assumptions.

Another non-material aspect that is relevant for post-growth capable models is time use. Analyses of time use can shed light on inequalities in non-monetised activities and in the invisibilised sphere of social reproduction (Mezzadri and Majumder, 2022; van Staveren, 2005). Furthermore, how individuals use their time may have strong effects on their wellbeing (Tomczyk et al., 2021).

## 5.2. Modelling transformation policies

Our framework of minimum requirements for modelling post-growth (Figure 4) is based on the elements identified as important for modelling *goal-oriented policy themes*. Modelling *transformation policy themes* requires additional elements and considerations. In this section, we therefore discuss four additional aspects that are prominent in the post-growth literature but not explicitly included in our survey-based minimum framework.

### 5.2.1. Different modes of provisioning

The provisioning systems framework has emerged as a way to understand how resources are transformed to satisfy human needs (Fanning et al., 2020; O'Neill et al., 2018). Different modes of provisioning can affect how resource use translates into need satisfaction (Hinton, 2021; Vogel et al., 2021). A distinction between different provisioning modes is crucial for modelling transitions towards economic democracy (Steinberger et al., 2024) or processes of decommodification (Fitzpatrick et al., 2022; Gerber and Gerber, 2017). Furthermore, the inclusion of non-institutional provisioning modes (e.g. within households or communities) could allow for the representation of non-monetised work and intersectional perspectives on inequalities within the provisioning process (Dengler and Plank, 2024; Power, 2004).

### 5.2.2. Disaggregated production and consumption

Another important issue is sectoral disaggregation, as the composition of production and consumption is crucial for assessing both social and environmental outcomes (Duchin, 1992; Gough, 2017; Hardt and O'Neill, 2017). A sufficient disaggregation of sectors or goods is important for reflecting differences in their environmental impacts and importance for basic need satisfaction (Baltruszewicz et al., 2021; Bärnthaler and Gough, 2023; Oswald et al., 2023). For instance, a disaggregation of different food types would allow for the modelling of a shift towards plant-based diets (Koch et al., 2017). Sectoral interdependencies can be modelled (e.g. through an input–output structure) to account for environmental impacts of infrastructure build-out or shifts in production, consumption, or end-use technology (Hickel et al., 2021; Slameršak et al., 2022).

### 5.2.3. International interactions

As national economies are open systems, an adequate model representation of national dynamics requires accounting for trade, international relations (e.g. debt, exchange rates), and cross-border financial flows (Hickel, 2017; Leoni et al., 2023). Ecologically unequal exchange can be assessed through the net-of-trade appropriation of biophysical resources and labour, which can be integrated into models using a multi-regional environmentally-extended input–output structure (Dorninger et al., 2021; Hickel et al., 2024). It is particularly important to understand how a post-growth transition in one country impacts — and is impacted by — other countries, highlighting the need to jointly model developments in several countries with different positions in the world system (Leoni et al., 2023). Furthermore, the study of international power structures and rules of global governance can help researchers understand their role in the creation of growth imperatives, both in the Global North and Global South (Lang, 2024).

#### 5.2.4. The financial system

Finally, although the financial sector was not ranked highly by respondents in our survey, the financial system affects the real economy, warranting the inclusion of the financial system in post-growth models (Jackson et al., 2014; Keen, 2011). Climate change may have severe effects on financial stability (Campiglio et al., 2018; Dafermos et al., 2018), and the financial sector plays an essential role in transitioning towards a more sustainable economy (Antal and Van Den Bergh, 2013; Victor and Jackson, 2020). Furthermore, post-growth scenarios may substantially increase public debt-to-GDP ratios or drive inflation, which could provoke the central bank to raise interest rates. They could also affect private debt-to-GDP ratios, a key indicator of financial stability (Keen, 2014). This highlights the need to assess economic stability under post-growth scenarios (Edwards et al., 2025). A key approach for modelling monetary dynamics and endogenous money is stock–flow consistent accounting (Godley and Lavoie, 2012). Although several existing models in the post-growth literature adhere to the principles of stock–flow consistency, the representation of the financial system often remains relatively simple (Victor and Jackson, 2020). Therefore, a more detailed representation of financial actors and assets would be required to understand the opportunities and challenges in transforming the financial system.

# 6. Discussion

One key contribution of this research is the conceptual framework that outlines minimum requirements for a post-growth model and how these requirements could be integrated. We acknowledge that probably no existing model meets the full list of requirements that we have identified, and that including all these requirements involves a major effort. Our aim is not to criticise existing models but to bring structure to the debates on modelling post-growth scenarios and to offer an aspirational goal for researchers working on these topics, while also providing a framework for evaluating attempts to model post-growth.

We suggest that the contributions of this research can be used in four ways. First, the framework can be used to assess how adequate existing models and modelling studies are for simulating or assessing post-growth scenarios. Second, it can be used as a guide for modellers from various communities to extend existing models to bring them more in line with the post-growth literature. Third, modellers interested in specific policy themes or research questions can use the results related to specific policy themes to identify important elements that should be included. Lastly, the framework can introduce newcomers to the main concepts that are relevant for post-growth modelling.

We now discuss tensions between the survey results and the post-growth literature, limitations of modelling for exploring post-growth futures, and limitations of this study.

## 6.1. Tensions between the survey results and the post-growth literature

The first tension relates to the modelling of the financial system. Several authors have argued for the importance of representing the financial system in models (e.g. Antal and Van Den Bergh, 2013; Campiglio et al., 2018; Victor and Jackson, 2020). Furthermore, heterodox macroeconomics has stressed that the current finance-led globalised capitalist system shapes many aspects of society and economy, such as firm incentives, labour markets, and public budget and tax systems (Aglietta, 2000; Boyer, 2000; Lazonick and O'Sullivan, 2000). However, the political economy of financialised capitalism is often not or poorly represented in post-growth models. Additionally, the survey results show that respondents do not consider the inclusion of the financial system as directly relevant for modelling goal-oriented policy themes. This discrepancy may partially be explained by the overall relatively low expertise on the financial system amongst our survey respondents (Supplementary Figure S1). Notwithstanding this lack of expertise, modellers should aim to better represent how the current global political economy could present barriers to a post-growth transition.

The inclusion of a diverse range of environmental feedback mechanisms and their interactions is another important gap in the post-growth modelling literature. This gap is reflected in the survey results, as there was no consensus between the respondents that ecological impacts are required to model any of the policy themes. The high uncertainties relating to the magnitude and channels of feedback mechanisms make it challenging to model them (Van Eynde et al., 2024). Even for climate damage functions — one of the most studied environmental feedback mechanisms — the estimated effect on long-term GDP growth varies substantially between studies (Morris et al., 2025). Nevertheless, these feedback mechanisms should form an integral part of post-growth models, as changes to Earth systems will have increasing impacts on human wellbeing (e.g. Kaczan and Orgill-Meyer, 2020; Pecl et al., 2017).

Finally, the results showed no consensus that non-institutional provisioning should be included for any of the policy themes. Feminist scholars have pointed out that the representation of social provisioning is incomplete if essential, non-monetised parts are not considered (Dengler and Plank, 2024; Power, 2004). Forms of provisioning that fall outside of formal institutions can be an important strategy for social transformation and resistance to capitalism (Caffentzis and Federici, 2014; Gibson-Graham et al., 2016). Furthermore, decommodification of provisioning would have negative effects in a model that only considers monetised flows, as the non-market provisioning would “leak” out of the model. Although some work has been undertaken in models of different sizes to represent care and the informal household economy (Blecker and Braunstein, 2022; Kemp-Benedict, 2025), the modelling of non-institutional modes of provisioning remains an important gap in post-growth modelling research.

### 6.2. Limitations of modelling for exploring post-growth futures

Modelling is an important tool for understanding systemic effects of post-growth scenarios, but it is subject to limitations. A first challenge is the modelling of structural changes that the post-growth literature proposes (Edwards et al., 2025). Because most macroeconomic models build at least partly on econometric estimations of relationships based on past data, they may become invalid in a future where these relationships change qualitatively, as one may expect from structural change (Lucas Jr, 1976). Moreover, existing macroeconomic theories that describe the current socioeconomic system may not be suited to describe the radically different systems that a post-growth transition aims to achieve.

A second challenge relates to the limits of what mathematical models can represent. Although models should include more intangible constructs (e.g. social connection) that matter for human wellbeing (Van Eynde et al., 2024), there are limitations to the insights that macroeconomic models can provide. This type of modelling should thus be used as complementary to, and within a structured plurality of, different methods that can be used to further our understanding of how to achieve post-growth goals (Berik and Kongar, 2021, chap. 13; Spash, 2012).

Lastly, model results are affected by the underlying pre-analytic vision and assumptions of those who develop the model (Meadows and Wright, 2008, p. 163; Sgouridis et al., 2022). Furthermore, the economic theory upon which a particular model is based can significantly affect the outcomes of modelled transition scenarios (Mercure et al., 2019). Due to the size of most models used in post-growth scenario research, it is challenging to understand how model outcomes are affected by embedded assumptions. One way to mitigate this is to perform model intercomparisons where the same scenarios are simulated in different models of the same system and assessed on the same outcome variables, as has been done in the context of IPCC assessments (O'Neill et al., 2016; Zelinka et al., 2020). Furthermore, researchers can innovate and test new ideas in small models before including them in more complex models.

### 6.3. Limitations of this research

This study has several limitations. First, some respondents found it difficult to determine when there was a direct link between an element and a policy, which calls for caution in interpreting the results of the survey. However, by basing our findings on the elements where there is a statistically significant consensus, we can at least highlight the agreement within the community. Although we propose a set of minimum requirements, we don't claim that this list is exhaustive. The actual requirements and design for any model depends on its scope and research question (Victor and Jackson, 2020).

Second, as we take a positivist approach by basing a normative framework on the expertise and agreement within the community, the results may be skewed towards dominant views and topics that are already well understood. Furthermore, our sample is relatively small and consists of people who are working on models or who have published studies on modelling. Consequently, the responses may be affected by what is contained in current models and may not adequately reflect what should be included to advance post-growth modelling.

Third, we used the survey responses related to the goal-oriented policy themes to determine which elements to include in our framework for a post-growth model. We chose this approach because these themes relate closely to the three post-growth goals, which in our view are defining parts of the post-growth literature. However, other criteria could have been used to decide which elements should be included in the framework.

## 7. Conclusion

As interest in modelling post-growth is increasing, clarity is needed on what is required for models to adequately represent and assess post-growth scenarios. So long as these requirements are not established, post-growth ideas may be misrepresented in scenario modelling exercises, for instance by conflating post-growth scenarios with recessions. To address this issue, this article proposes a conceptual framework of minimum requirements for a post-growth model.

At a high level, we argue that models should be able to evaluate scenarios in terms of environmental sustainability, social equity, and human wellbeing, while reflecting the post-growth literature's critiques of the pursuit of economic growth. Doing so requires representing the biophysical, economic, and social spheres and their interdependencies.

To represent the biophysical sphere, models should include environmental limits, represent feedback mechanisms between society, economy, and the environment, treat energy as a factor of production, and express resource and pollution flows in biophysical units. In the economic sphere, households should be disaggregated, with income and wealth tracked per group. For firms, there should be a representation of technological change and resource/pollution intensities. Key government interventions related to redistribution, access to essential goods, and environmental protection should be included. Furthermore, models should represent production and consumption dynamics and calculations of GDP endogenously and acknowledge limits to decoupling. Regarding the social sphere, material and non-material need satisfiers, and the satiability of needs should be represented, as should time use and the affordability of essential goods and services. Moreover, GDP should not be treated as a proxy for wellbeing and should only be used as a driver of wellbeing indicators if that is theoretically and empirically supported.

Other elements, such as modes of provisioning, sectoral disaggregation, international interactions, and the financial system, are also relevant for modelling transformation policy themes or addressing specific research questions. Important research gaps remain around the understanding and modelling of the financial system, environmental feedback mechanisms, and non-institutional modes of provisioning.

This research contributes to the emerging field of post-growth modelling and aims to facilitate the uptake of post-growth scenario modelling in various modelling communities. Our framework can be used as a tool to assess and extend existing models, to inform the development of new models, and to communicate core ideas from the post-growth literature to other modelling communities. Given the challenges of including the full list of requirements in a model, the framework can be seen as a guide towards an aspirational goal. By providing guidelines for aligning models with the normative and analytical foundations of the post-growth literature, this work enables the development of modelling tools that can adequately explore sustainable and just futures that go beyond economic growth.

## CRediT authorship contribution statement

**Reo Van Eynde**: Conceptualisation, Methodology, Software, Validation, Formal analysis, Investigation, Data Curation, Writing – Original Draft, Writing – Review & Editing, Visualisation

**Kevin J. Dillman**: Conceptualisation, Methodology, Investigation, Writing – Original Draft, Writing – Review & Editing

**Jefim Vogel**: Conceptualisation, Methodology, Investigation, Writing – Original Draft, Writing – Review & Editing

**Daniel W. O'Neill**: Conceptualisation, Methodology, Investigation, Writing – Review & Editing, Supervision, Project administration, Funding acquisition

## Acknowledgements

This research was funded by the European Union in the framework of the Horizon Europe Research and Innovation Programme under grant agreement numbers 101094211 (ToBe: "Towards a Sustainable Wellbeing Economy: Integrated Policies and Transformative Indicators"), and 101137914 (MAPS: "Models, Assessment, and Policies for Sustainability"). We want to thank Eric Kemp-Benedict and Andrew Jackson for their feedback on the survey design, and the two anonymous reviewers for their helpful comments. Furthermore, we thank the survey respondents for taking the time to fill in the survey.